# Test of High Time Resolution MRPC with Different Readout Modes

S. Yang[a,c], Y.J. Sun[a,c]*, C. Li[a,c]*, Y.K. Heng[b,c], S. Qian[b,c], H.F. Chen[a,c], T.X. Chen[a,c], H.L. Dai[b,c],

H.H. Fan[a,c], S.B. Liu[a,c], S.D. Liu[b,c], X.S. Jiang[b,c], M. Shao[a,c], Z.B. Tang[a,c], H. Zhang[a,c], Z.G. Zhao[a,c]

a Department of Modern Physics, University of Science and Technology of China(USTC), Hefei 230026, China

b Institute of High Energy Physics, Chinese Academy of Sciences(IHEP), Beijing 100049, China

c State Key Laboratory of Particle Detection and Electronics(USTC-IHEP),China

## Abstract

In order to further enhance the particle identification capability of the Beijing Spectrometer (BESIII), it is proposed to upgrade the current end-cap time-of-flight (eTOF) detector with multi-gap resistive plate chamber (MRPC). The prototypes, together with the front end electronics (FEE) and time digitizer (TDIG) module have been tested at the E3 line of Beijing Electron Positron Collider (BEPCII) to study the difference between the single and double-end readout MRPC designs. The time resolutions (sigma) of the single-end readout MRPC are 47/53 ps obtained by 600 MeV/$c$ proton/pion beam, while that of the double-end readout MRPC is 40 ps (proton beam). The efficiencies of three MRPC modules tested by both proton and pion beam are better than 98%. For the double-end readout MRPC, no incident position dependence is observed.

## 1. Introduction

The Beijing Spectrometer (BESIII) [1] is a high precision general-purpose detector designed for high luminosity $e^+e^-$ collisions in the τ-charm energy region at the Beijing Electron Positron Collider

sunday@ustc.edu.cn (Y.J. Sun)
licheng@ustc.edu.cn (C. Li)

(BEPCII) [2]. At BESIII, the current end-cap time-of-flight (eTOF) detector consists of 2×48 fast scintillators (EJ204) readout with fine-mesh photomultiplier tubes (Hamamatsu R5924) [3]. The time resolution (sigma) measured is 110 ps for muons in the di-μ events, 138 ps for pions (1 GeV/$c$), and 148 ps for electrons in the Bhabha events. The momentum range for K/π separation (greater than 2σ) is limited to 1.1 GeV/$c$ [4]. A GEANT4 simulation has been carried out [5] to study the cause of the degradation of time resolution for electrons. The major reason is found to be the multiple scattering effects upon materials (~0.28 $X_0$) between the main drift chamber (MDC) end-cap and the eTOF (eg. the MDC support structure, MDC readout electronics and cables). The produced secondary particles (mainly γ, electrons and positrons) cause a high multi-hit probability for electron events (around 71.5%) in each eTOF scintillator module, which makes the position-dependent time calibration difficult. A new particle detection technique insensitive to γ and a smaller readout cell size are thus required for the eTOF upgrade.

The multi-gap resistive plate chamber (MRPC) [6], with good time resolution, high detection efficiency [7,8] and relatively low cost, is a good choice. It can be produced with high granularity and is relatively insensitive to γ. Furthermore, the MRPCs have been widely used as TOF detector in many experiments, such as STAR at RHIC [9], ALICE at LHC [10], HADES and FOPI at GSI [11,12].

A proposal was raised in 2010 to upgrade the current BESIII eTOF with the MRPC technology, aiming at an overall 80 ps time resolution for minimum ionizing particles (MIPs). This will extend the momentum range for K/π separation (2σ) to 1.4 GeV/$c$. The overall 80 ps time resolution includes the uncertainty from bunch length (35 ps), uncertainty from clock system (~20 ps), uncertainty from θ-angle (~30 ps) and uncertainty from expected time of flight (30 ps) [1]. With these contributions subtracted, the intrinsic MRPC time resolution (including electronics) should thus be better than 55 ps. Three MRPC prototypes with different readout modes were manufactured and jointly tested with the front end electronics (FEE) and time digitizer (TDIG) modules dedicated to the eTOF upgrade in 2011 at the BEPC E3 line. The test results of timing and efficiency characteristics are presented in this paper, focusing on the comparison between the two readout modes. Such work offers an important reference for the final BESIII eTOF design.

This paper is organized as follows. Section II describes the MRPC modules. The readout electronics and beam test setup are presented in Sects. III and IV. Section V describes the performance of the MRPC modules including efficiency and time resolution. Section VI provides a concluding summary.

## 2. The MRPC modules

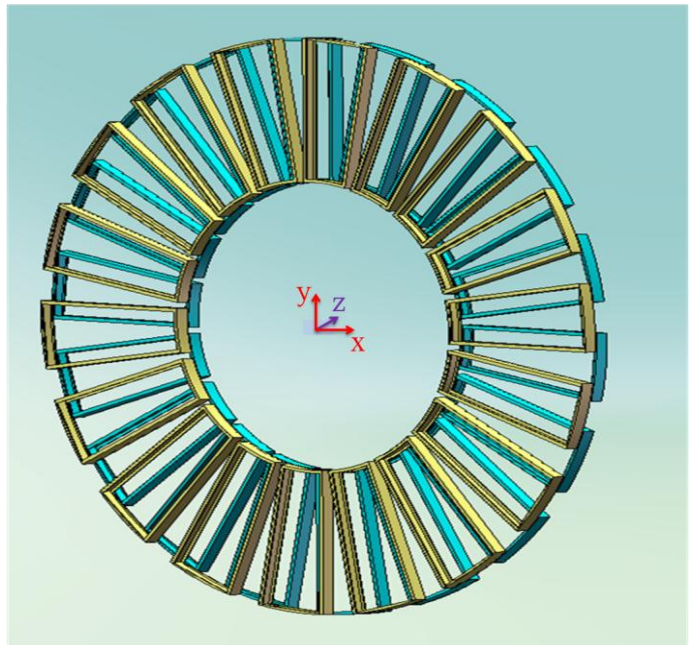

Fig. 1 The conceptual drawing of BESIII MRPC-based eTOF.

In the conceptual design, each BESIII end-cap ring [13] will be fully covered by 36 identical trapezium MRPC modules, as shown in Fig. 1. The effective inner radius of the ring is 478 mm while the outer radius is 822 mm. The new rings will be installed in the position of current eTOF, as shown in Fig. 2. The length of each MRPC module is 35.2 cm while the widths are shown in Fig. 3. Two readout modes, namely single-end and double-end, are considered. The single-end readout MRPC consists of 2×12 readout cells. The lengths of the cells range from 4.1 cm to 6.8 cm. The double-end readout MRPC has 12 readout cells with the lengths ranging from 8.6 cm to 14.1 cm. The interval between any adjacent cells is 4mm. Among the three modules tested, two have single-end readouts and one has double-end readouts. The estimated multi-hit probabilities from simulation [5] are 16.7% and 21.8% for the single-end and double-end readout MRPC, respectively.

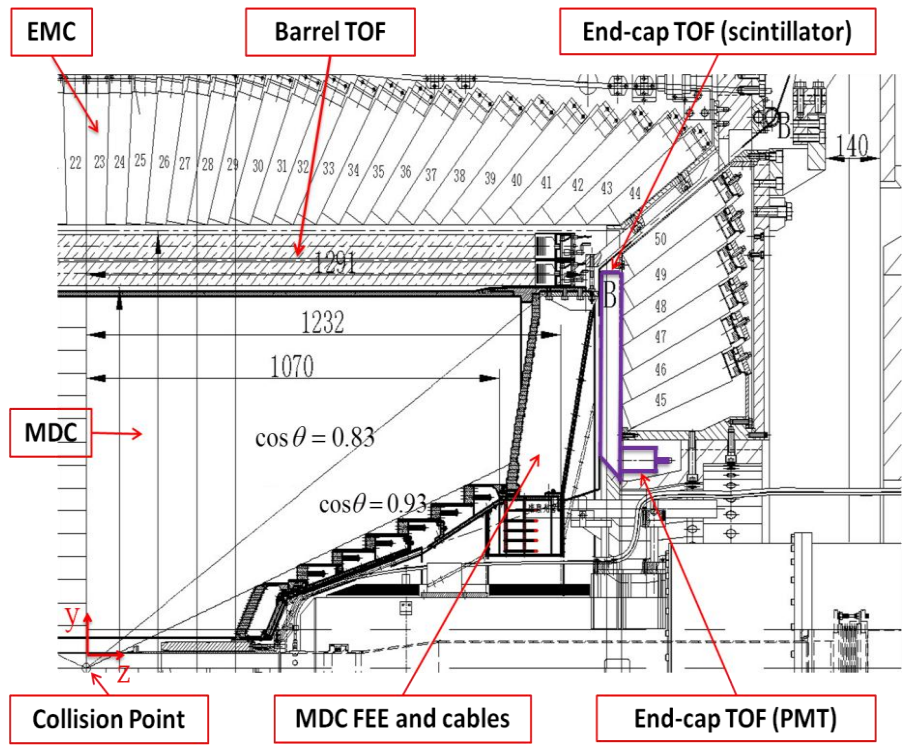

Fig. 2 The schematic of BESIII. The unit of length is mm.

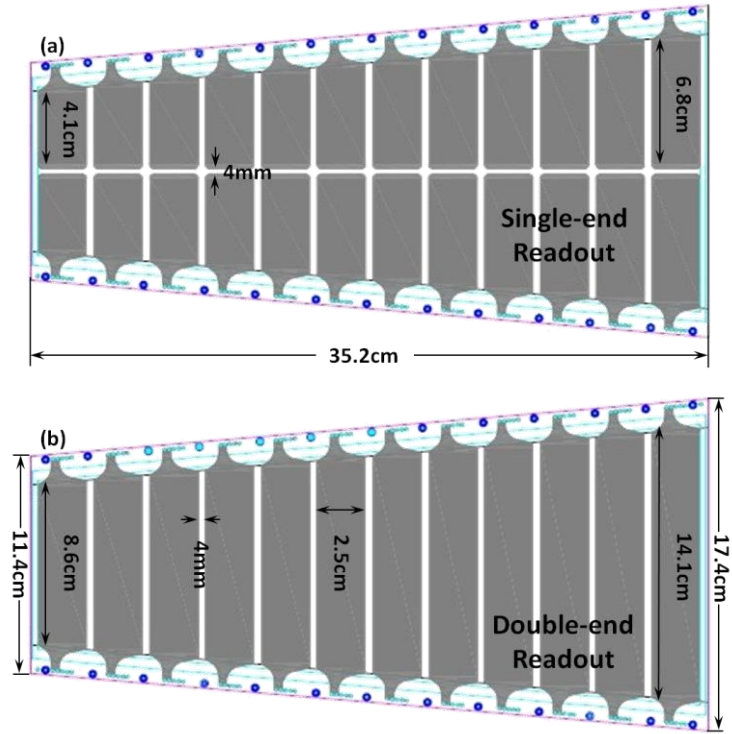

Fig.3 The layouts of two printed circuit boards (PCB) of the two readout patterns (single-end readout (a) and double-end readout (b)) from a top view.

Fig. 4 [14] schematically shows a cross-sectional view of the MRPC prototypes. All the MRPC prototype modules have 12 gas gaps arranged in a double-stack configuration mirrored with respect to the central electrode. The gap width is 220 μm, defined by nylon fishing line. Floating glass sheets with volume resistance of ~$10^{13}$ Ω·cm are used as the resistive plates. The thicknesses are 0.4 mm and 0.55 mm for the inner and outer glass, respectively. The outer surfaces of the outermost glass in each stack are coated with graphite tapes, which serve as high voltage electrodes. The surface resistivity of the graphite tape is about 200 kΩ/□. Two pieces of 3 mm thick honeycomb-board are attached to the outer surfaces of the detector to reduce structural deformations. The MRPC prototype module is placed in a gas-tight aluminum box whose total thickness is 2.5mm (~0.028 $X_0$), flushed with a standard gas mixture (90% Freon + 5% $SF_6$ + 5% iso-$C_4H_{10}$).

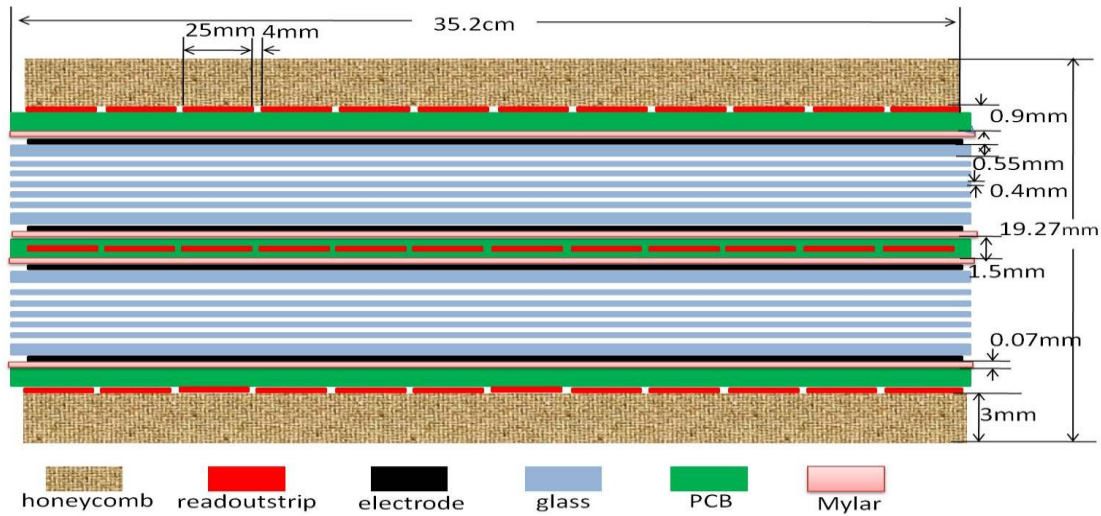

Fig.4 The schematic drawing of the cross-section of the MRPC

## 3. The readout electronics

The FEE makes use of the NINO chip developed by the ALICE-TOF group [15]. Each FEE module features 24 differential input channels and outputs correspondent LVDS signal with the signal charge encoded in its width. The timing accuracy (RMS) can be better than 15 ps for each channel when the input charge is larger than 100 fC [16]. The FEE board is fixed on the surface of the aluminum gas box which contains the MRPC module in order to reduce the input capacitance. A flexible printed circuit is designed to connect the MRPC module output with proper impedance (54 ohm). The connector (VRDPC-68-01-M-RA) with 86 pins and the shielded differential cable (VPSTP-24-5000-01 from SAMTEC) are used to connect the FEE and the TDIG [17], aiming at reducing the time jitter from signal transmission and ensuring the signal quality.

The TDIG modules, relying on the ASIC HPTDC chip developed by the microelectronics group at CERN [18,19], focus on receiving and digitizing the signals from the FEE, packing the data with predefined format and uploading them to the data acquisition (DAQ) system via the VME bus. Each TDIG board integrates 72 channels with 9 HPTDC chips operating in the high resolution mode. The time resolution of the TDIG module can achieve 20ps (RMS) after an integral non-linearity (INL) compensation [17].

The coincidence threshold test power (CTTP) module provides power, threshold and test signals to the FEE. It also receives the OR differential signals from the FEE and produces fast hit signals after coincidence. The hit signals are used by the TOF trigger subsystem [20]. The schematic of the readout electronics system for the eTOF upgrade is shown in Fig. 5, including 72 FEE, 24 TDIG and 2 CTTP modules. The TDIG and CTTP modules are housed in two VME64xP crates, which are responsible for the two eTOF rings. In this beam test, only one FEE and one TDIG module are used.

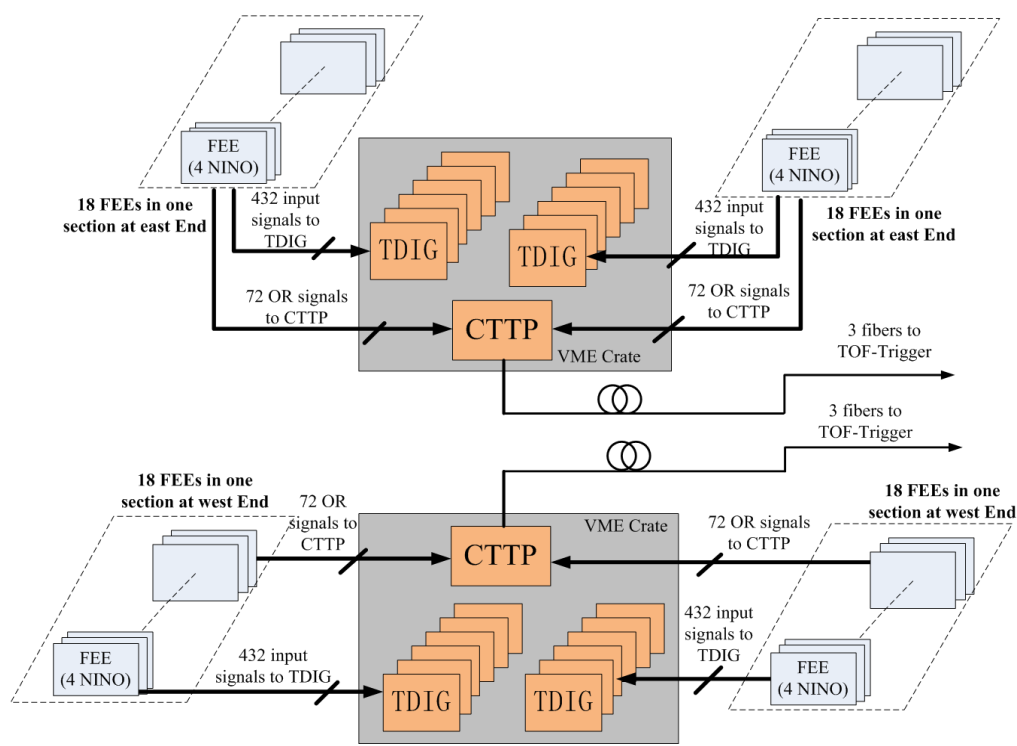

Fig. 5 The schematic of readout electronics of the whole MRPC-based eTOF system

## 4. The beam test setup

A beam test was performed at the E3 line of BEPCII using the secondary particles (mainly $e^{+/-}$, $\pi^{+/-}$, p) from an incident electron beam hitting a carbon target [21]. The momenta of the secondary particles are around 600 MeV/$c$. Among these secondary particles, protons are dominant. Since the available beam time was limited, the statistics is low for pion events. Consequently the results below are mainly from proton events.

The setup of the beam test is shown in Fig. 6. The Cherenkov detector (C0) is used to veto the electron. The MRPC module, placed in a gas-tight aluminum box, is fixed on a movable platform. The differential output signals of the MRPC are fed to the FEE directly. The coincidence signal of two larger scintillators (S1, S2) and four smaller scintillators (T1-4) is used as the trigger of the beam test DAQ system. Meanwhile, the four smaller scintillators ($2 \times 5$ cm$^2$ active area) both provide a reference time ($T_0$) for the MRPC module, and identify the incident particle species through their charge spectra, as shown in Fig. 7. The signals from the $T_0$, after discrimination and LVDS conversion, are sent to the TDIG for a leading-edge time measurement. The charges of the $T_0$ signals are measured by a charge-to-digital converter (QDC) module after proper delay. The logic diagram of the beam test DAQ system is shown in Fig. 8.

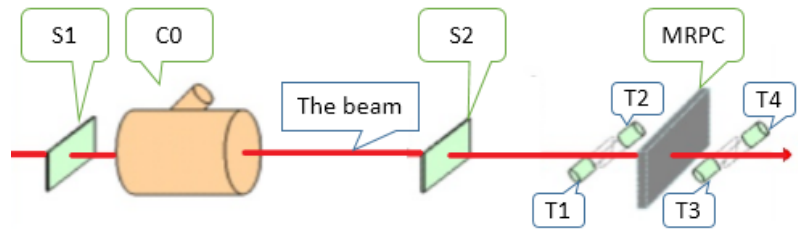

Fig. 6 The setup of beam test experiment.

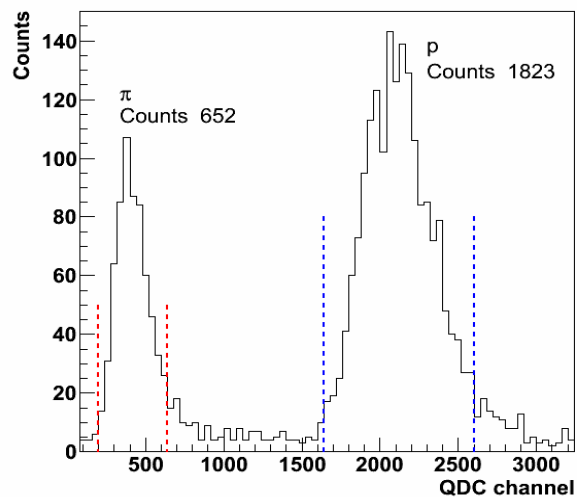

Fig. 7 Charge spectrum of PMT4 (T4 in Fig.6). Pion and proton are well separated.

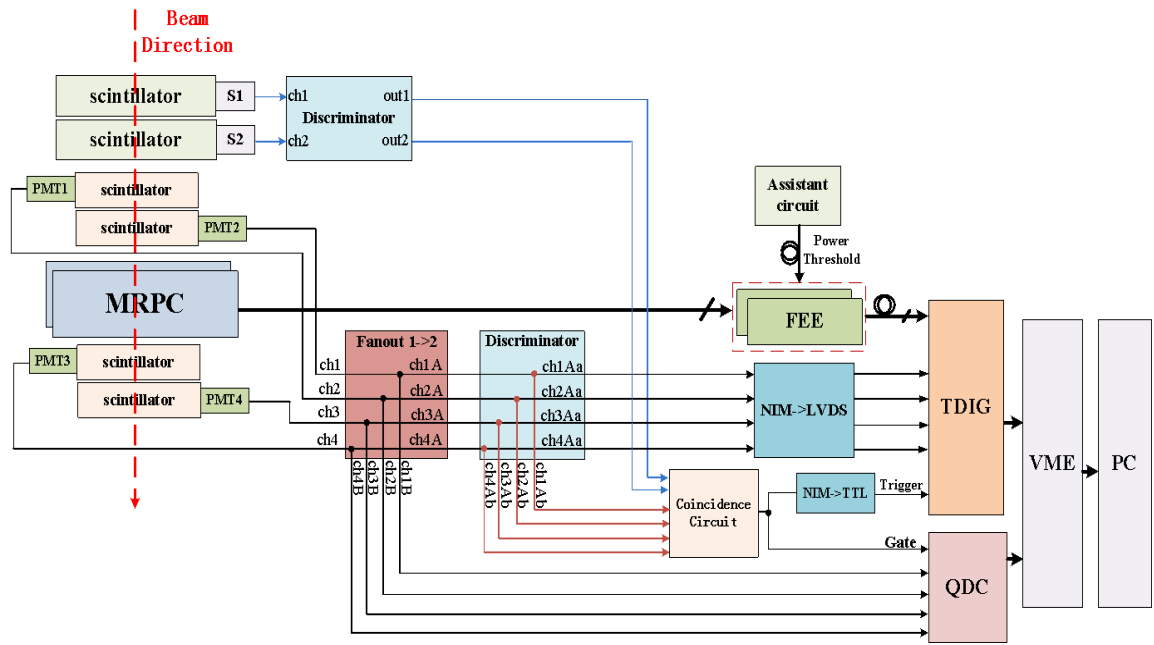

Fig. 8 The logic diagram of the beam test DAQ system.

# 5. The performance of MRPC modules

In order to find the intrinsic time resolution of the MRPC module, the jitter of $T_0$ should be determined first. $T_0$ is defined as the average timing of the four PMTs $(t_1+t_2+t_3+t_4)/4$. A typical $T_0$ time resolution after the slewing correction is 29 ps, as shown in Fig. 9. The time measured by the MRPC is also corrected for the slewing effect.

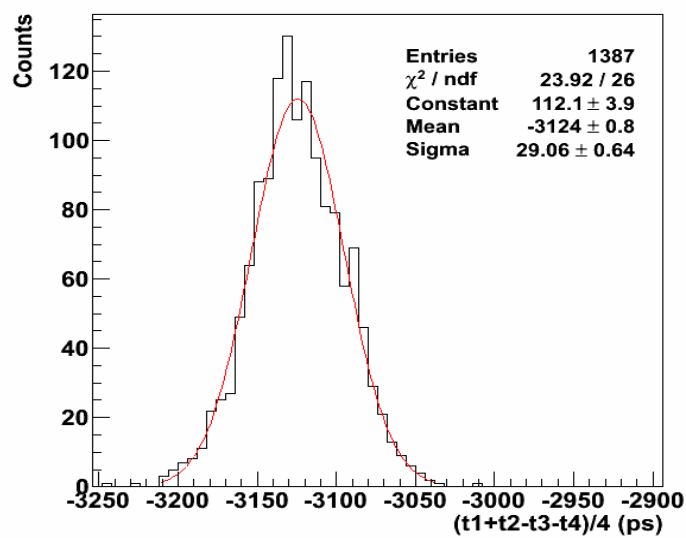

Fig. 9 The distribution of (t1+t2-t3-t4)/4 ($\sigma$=29 ps) after the slewing correction.

Fig. 10 shows the time-over-threshold (TOT) spectrum of the MRPC, the time-TOT correlation and the MRPC time distribution before and after the slewing correction. The overall time resolution after slewing correction is 52 ps. With the start time resolution 29 ps subtracted, the MRPC intrinsic time resolution is 43 ps.

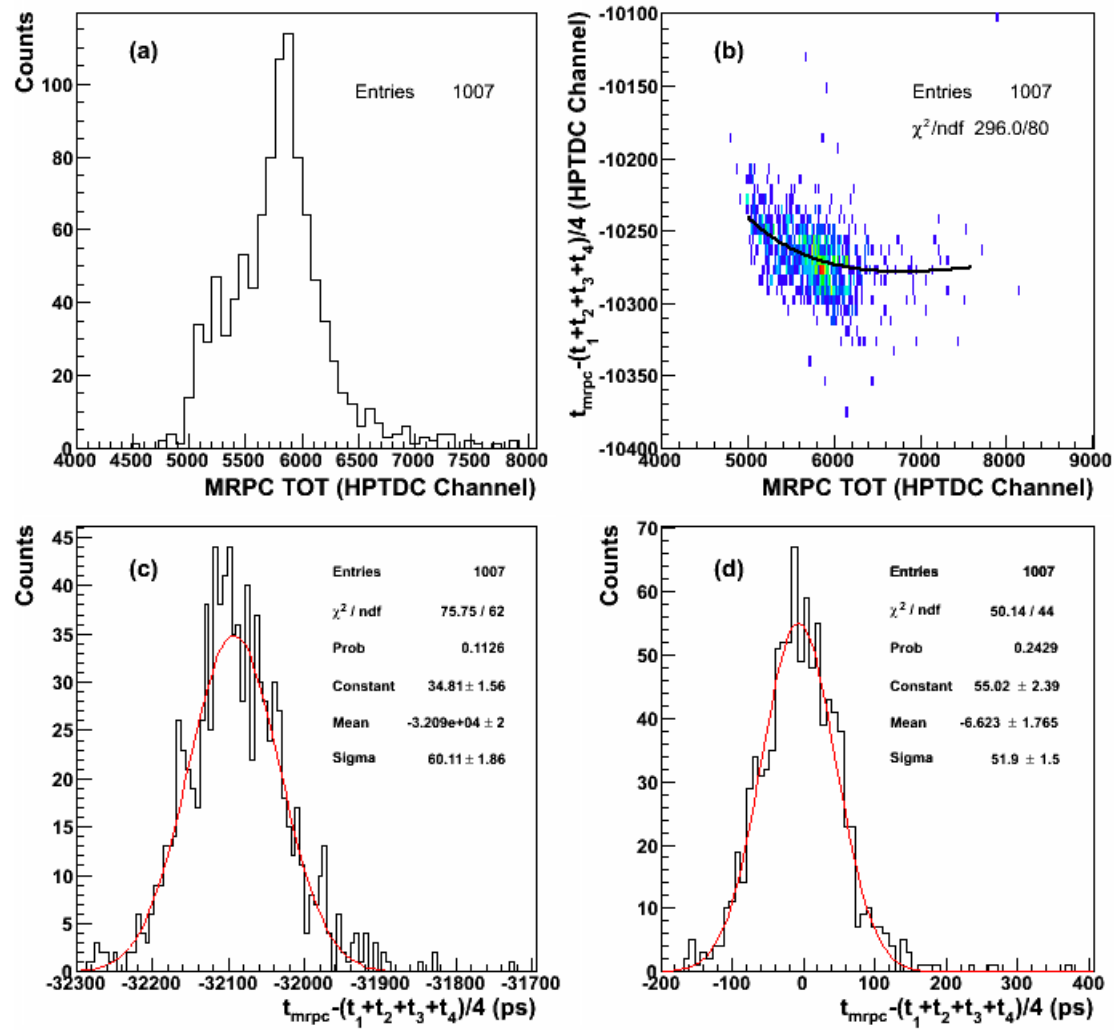

Fig. 10 (a) TOT spectrum of MRPC at operating voltage ; (b) MRPC raw time-TOT correlation and the fitted correction curve; (c) MRPC time distribution before the slewing correction; (d) MRPC time distribution after the slewing correction.

## 5.1 The HV scan

The efficiency and time resolution are scanned as a function of the high voltage (HV) in order to find the optimum operation voltage of the MRPC. The MRPC efficiency plateau and time resolution of the two different readout modes are shown in Fig. 11. For the single-end readout mode, the detection efficiency is above 98% when the applied HV is higher than ±5.8 kV for protons and ±6.6 kV for pions (MIP). To keep the efficiency of the double-end readout MRPC above 98%, the applied HV has to be greater than ±6.4 kV and ±7.0 kV for protons and pions, respectively.

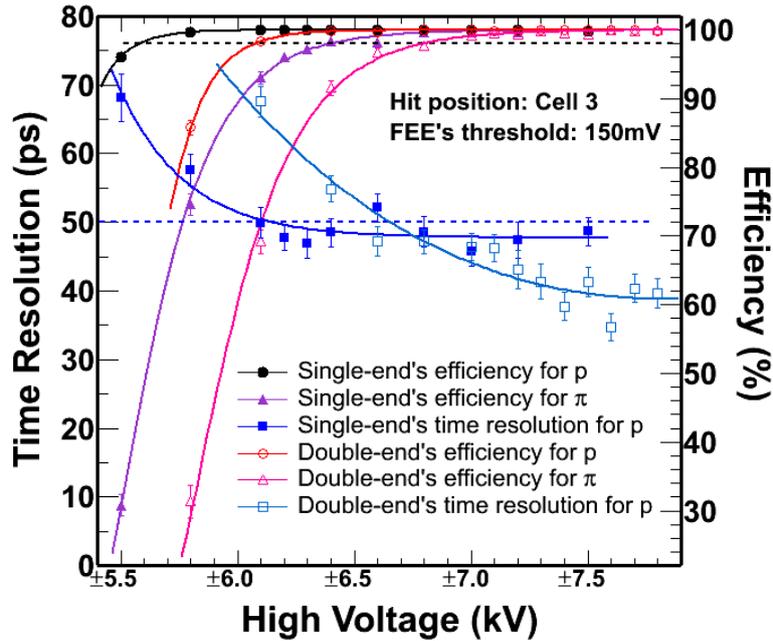

Fig. 11 The HV dependence of the efficiency and time resolution of two different readout MRPCs.

Comparing the two different readout modes, it is clear that the double-end readout MRPC needs higher working HV to achieve better time resolution. This is understood since each end of the double-end readout MRPC cell shares the induced signal charge almost equally. For the results reported in Sect. 5.2, the corresponding working HVs are ±7.0 kV and ±7.2 kV for the single-end and double-end readout MRPC, respectively.

In addition, for the single-end readout mode, we have accumulated enough pion events at three HV points. The effective number of pion events for each HV point is around 900. The time resolution of the single-end readout MRPC for pions is 53±3 ps, as shown in Fig. 12.

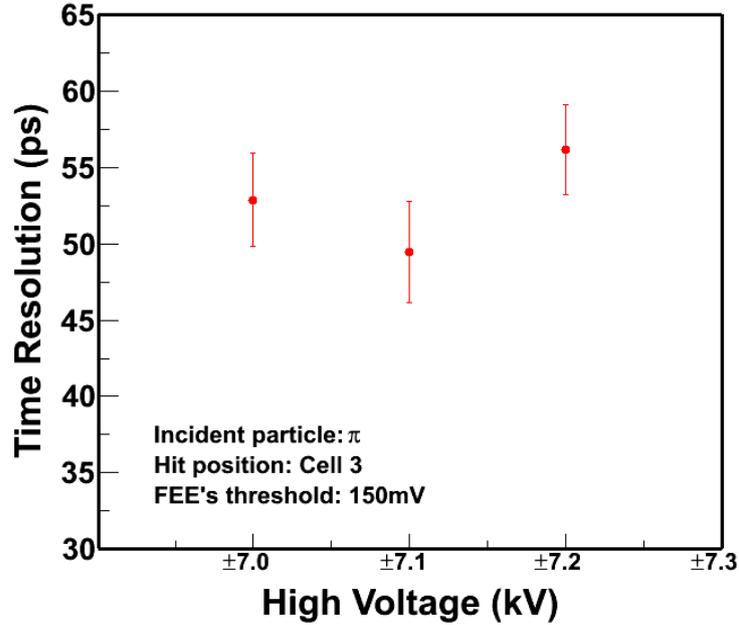

Fig. 12 Time resolution of single-end readout MRPC for pion events.

It should be noted that, for the time resolution of the double-end readout MRPC, the average time $((T_{end1}+T_{end2})/2)$ is independent of the hit position along the cell. This is not the case for the single-end readout MRPC. Since there is no tracking detector for the test beam, the position information comes only from the size (2 cm long) of the trigger detectors placed along with the readout cell. According to our previous measurements [22], the signal propagation velocity along the cell is around 0.02 cm/ps. The time jitter caused by the position uncertainty is thus estimated to be about 30 ps (RMS). This additional contribution should be subtracted from the time resolution of the single-end readout MRPC.

## 5.2 The beam position scan

### 5.2.1 The beam position scan across the cells

Since the MRPC modules are made in trapezium shape, the length of each cell varies. Scans across the cells are done to investigate the influence of different cell lengths. The MRPC modules are placed on a movable platform and the center of different cell can be moved to the trigger region. In the following discussion, the shorter cell starts with the lower cell ID.

For the single-end readout mode, two identical MRPC modules were placed close to each other and tested at the same time. The one upstream is defined as MRPC1, the other one is defined as MRPC2. They have similar performance as shown in Fig. 13. The time resolution is less than 50 ps for most cells except for the two outmost ones.

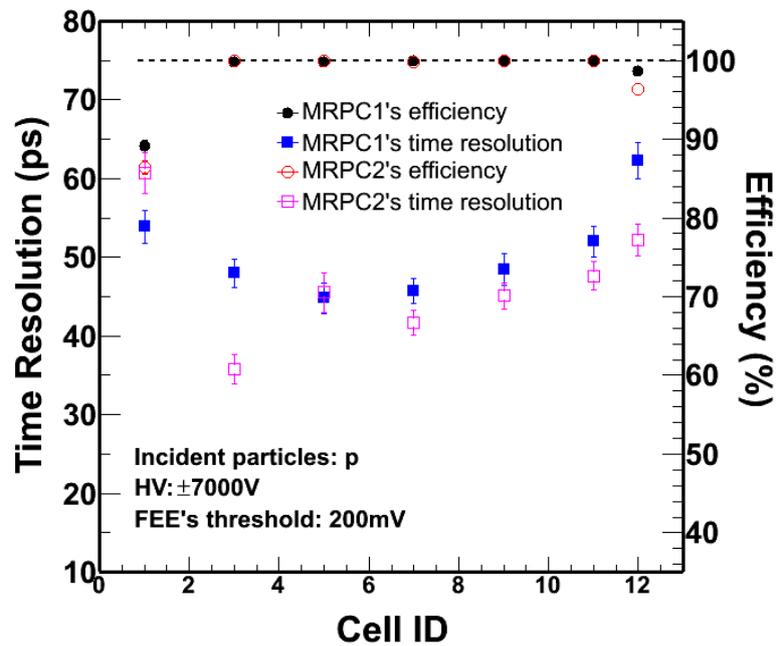

Fig. 13 The cross-cell scan result of the single-end readout MRPC.

Similarly, the double-end readout MRPC has also been scanned across the cells. The overall time resolution is ~40 ps, and the cell length has negligible influence on the time resolution, as shown in Fig. 14.

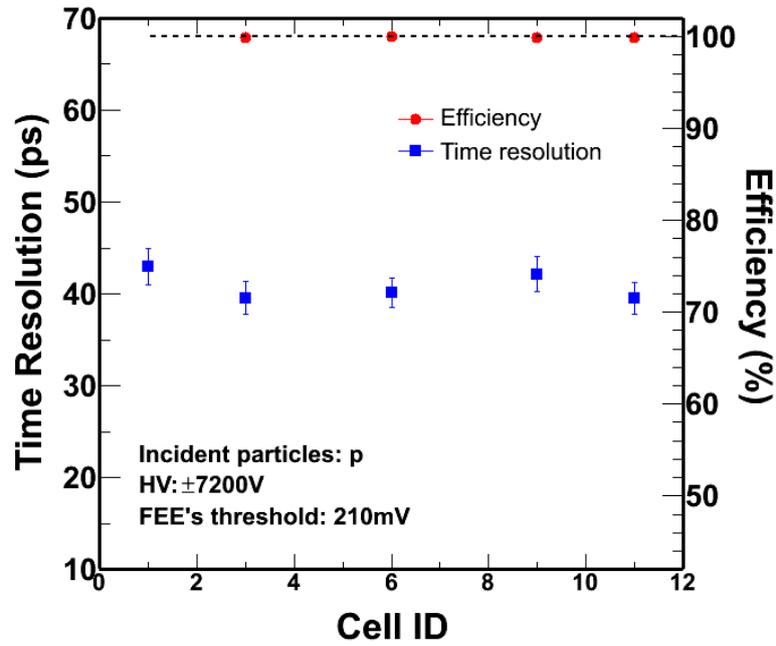

Fig. 14 The cross-cell scan result of the double-end readout MRPC.

### 5.2.2 The beam position scan along the cells

To investigate the detector performance with respect to the different particle incident position, especially for the single-end readout mode, a scan has been performed along the cell. Two cells (cell 6 and cell 11) are scanned with 1-cm step. As shown in Fig.15 and Fig. 16, the time resolution seems to be slightly worse when the hits are close to the readout end of the cell where signal is fed out, as previously observed in similar studies [23]. Further investigation with more precise tracking is foreseen to clarify it.

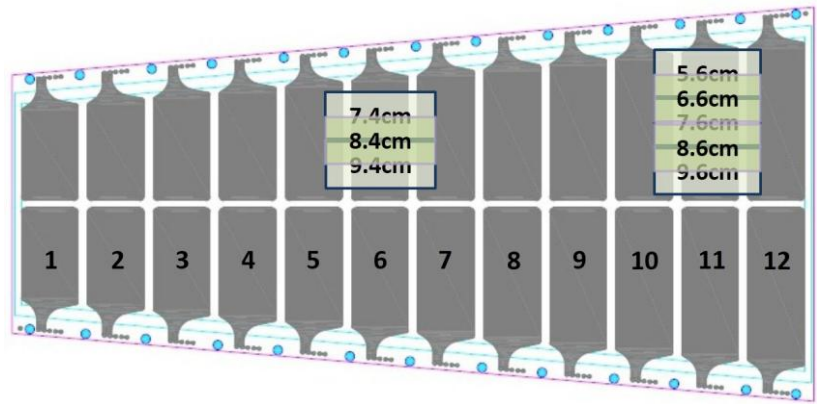

Fig. 15 The beam scan positions along the cells of the single-end readout MRPC, with a scan step of 1 cm.

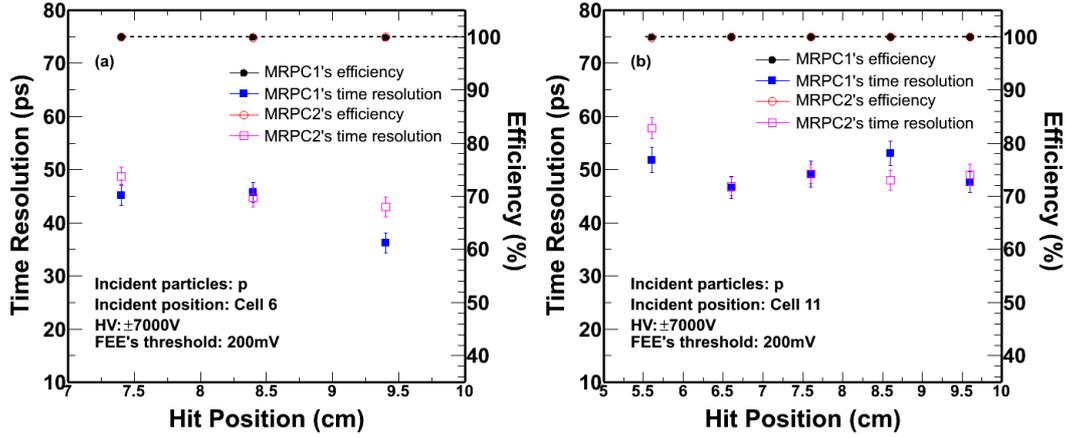

Fig. 16 Scans along the cells of the single-end readout MRPC for (a) cell 6 and (b) cell 11.

## 6. Conclusions

The MRPC prototypes with double-end and single-end readout cells, including the dedicated electronics, have been manufactured and tested for the BESIII eTOF upgrade. The results of the beam test at the BEPC E3 line prove the excellent performance of these MRPC modules. The efficiencies of all three MRPC prototype modules are higher than 98%. The time resolution of the double-end readout MRPC can reach 40 ps for 600 MeV/$c$ protons. After subtracting the contribution from the beam position uncertainty, the intrinsic time resolution of the single-end readout MRPC is expected to be better than 50 ps for pions (MIP). For the double-end readout mode, the incident position has negligible effect on the MRPC performance.

According to the results of this beam test, the performance of the double-end readout MRPC is insensitive to the position of incidence. Since the tracking performance is limited in precision within the BESIII eTOF acceptance [5], the double-end readout MRPC is a better choice for the BESIII eTOF upgrade according to this beam test.

## Acknowledgments

This work is supported by the National Natural Science Foundation

of China under Grant No. 10979003, 11275196, U1232206. We express our gratitude to Dr. Lijuan Ruan for her nice and constructive suggestions on the paper revision process. We would like to acknowledge Prof. Jiacai Li, Zunjian Ke, Dr. Guangpeng An and all members of the Beam Group of IHEP for their great support in the beam test.